\documentclass[%
reprint,
superscriptaddress,
 amsmath,amssymb,
aps,
pra,
longbibliography
]{revtex4-2}

\usepackage{graphicx}
\usepackage{ulem}
\usepackage{xcolor}
\usepackage{dcolumn}
\usepackage{bm}
\usepackage[mathlines]{lineno}
\usepackage{circledsteps}
\usepackage{hyperref}

\newcommand{\Rb}{$^{87}\text{Rb}$ }
\newcommand{\Cs}{$^{133}\text{Cs}$ }
\newcommand{\ket}[1]{\left\lvert #1 \right\rangle}
\newcommand{\abs}[1]{\left|#1\right|}

\begin{document}

\preprint{APS/123-QED}

\title{Enhancing the power of a  quantum heat engine via control of the\\ system--reservoir coupling}

\author{Sabrina Burgardt}
  \affiliation{Department of Physics and State Research Center OPTIMAS, RPTU University Kaiserslautern-Landau, D-67663 Kaiserslautern, Germany}
\author{Julian Feß}
  \affiliation{Department of Physics and State Research Center OPTIMAS, RPTU University Kaiserslautern-Landau, D-67663 Kaiserslautern, Germany}
\author{Silvia Hiebel}
  \affiliation{Department of Physics and State Research Center OPTIMAS, RPTU University Kaiserslautern-Landau, D-67663 Kaiserslautern, Germany}
\author{Eric Lutz}
  \affiliation{Institute for Theoretical Physics I, University of Stuttgart, D-70550 Stuttgart, Germany}
\author{Artur Widera}
\email[Contact author; E-mail: ]{widera@rptu.de}
  \affiliation{Department of Physics and State Research Center OPTIMAS, RPTU University Kaiserslautern-Landau, D-67663 Kaiserslautern, Germany}
\email{widera@rptu.de}

\date{\today}

\begin{abstract}
The non-equilibrium properties of open quantum systems are determined by the microscopic laws governing energy exchange with their environment. 
In particular, an enhancement of the performance of quantum heat engines has been predicted by speeding up the dynamics through control of the system--bath interaction. However, direct microscopic control of heat transfer between the machine and the reservoir has remained elusive so far. 
Here, we experimentally demonstrate such control in a quantum Otto engine realized with ultracold \Cs atoms coupled to an atomic reservoir of ultracold \Rb atoms.
Heat exchange between the two is mediated by inelastic $s$-wave collisions whose energy-dependent scattering cross sections lead to an asymmetric equilibration dynamics in the isochoric heating and cooling strokes.
By tuning the kinetic temperature of the atomic reservoir, we modify the associated microscopic scattering rates, and thereby the heat transfer law, giving control over the time allocation within the engine cycle through control over the microscopic, multi-exponential relaxation dynamics.
This enables power output optimization at fixed efficiency. 
Our results establish microscopic control of system-reservoir interactions as a tool for manipulating heat flow at the nanoscale and engineering the finite-time performance of quantum thermal machines.
\end{abstract}

\maketitle

\section{\label{sec:introduction}Introduction}
The performance of heat engines operating at finite time is fundamentally constrained by the interplay between energy conversion and dissipation. 
While reversible engines achieve maximal efficiency, they deliver vanishing power.
Optimizing the power output, therefore, constitutes a central problem in finite-time thermodynamics, both in the classical and quantum regimes~\cite{Curzon1975, Andresen1984,gor91,and11,kos13,Chen1994_carnot,bro05,sch08,esp09,Esposito2010,whi14}.
In the past decades, various strategies have been developed to speed up thermodynamic processes by external control, including shortcut-to-adiabaticity techniques, that allow the engineering of adiabatic dynamics in finite time by suppressing nonadiabatic excitations~\cite{tor13,gue19,hou25}, and shortcut-to-thermalization schemes, that accelerate the relaxation of a system towards its equilibrium state~\cite{mar16,ray23,dan19}. 
These methods decrease the cycle time and, thus, increase the power output of a thermal machine. They are usually implemented by adding time-dependent terms to the system Hamiltonian, and generally come with additional work or dissipation costs~\cite{tor13,gue19,hou25,mar16,ray23,dan19}. 
Recently, an alternative approach to enhance engine performance based on the modulation of the system-bath coupling has been put forward~\cite{pan20}. 
It has been suggested that such speed-up protocols might be realized without increasing overall dissipation~\cite{pan20}. 
However, controlling system--bath interactions is difficult in most systems.

At the same time, the coupling between system and reservoir significantly impacts the heat transfer between the two. 
By setting dynamical constraints on energy exchange, the specific form of the heat transfer law determines the optimal operating regime of a thermal machine~\cite{Gordon1990, Chen1994, Feldmann1996, Deffner2018}. 
However, most existing theoretical studies rely on phenomenological (macroscopic) descriptions of heat transfer~\cite{Curzon1975, Andresen1984,gor91,and11,kos13,Chen1994_carnot,bro05,sch08,esp09,Esposito2010,whi14}, for example, Newton's linear law of cooling which corresponds to an exponential heat relaxation behavior. 
While microscopic heat engines have been realized across a variety of experimental platforms, including trapped ions~\cite{Rossnagel2016, Lindenfels2019,van20}, molecules~\cite{Volosheniuk2025}, nuclear magnetic resonance~\cite{pet19,ass19}, solid-state systems~\cite{jos18,Klatzow2019, Kwon2025}, and ultracold atomic setups~\cite{Brantut2013,Bouton2021,Nettersheim2022,kim22,Koch2023}, experimental control over the underlying heat transfer mechanism \mbox{itself -- particularly} over its dependence on \mbox{microscopic parameters -- has} remained limited. 
Achieving such microscopic control is not only relevant for the optimization of thermal machines, but is also of fundamental interest for the understanding and manipulation of heat flows at the nanoscale~\cite{dub11,li12,pek21}.
More generally, understanding how microscopic interaction processes determine macroscopic relaxation dynamics is a central theme across many areas of quantum physics, ranging from collision-model descriptions of thermalization~\cite{Ciccarello2022}, to spin-bath systems~\cite{Prokofev2000} and engineered open quantum systems~\cite{har22}. 
A common feature of these systems is that equilibration is not necessarily governed by a single relaxation time. 
Understanding how this relaxation spectrum affects finite-time performance remains an outstanding challenge.

In this article, we demonstrate experimental control over the heat transfer law of a quantum heat engine and exploit it to optimize the power output.
Specifically, we realize a quantum Otto heat engine~\cite{kos17} driven by atomic collisions, in which the system-reservoir coupling is governed by energy-dependent $s$-wave scattering processes~\cite{coh11}. 
The working medium consists of the seven Zeeman states of the internal hyperfine ground-state manifold of individual \Cs atoms. 
These atoms are immersed in an ultracold thermal cloud of \mbox{\Rb atoms}, which serves as an atomic spin reservoir~\cite{Prokofev2000}.
The two atomic species interact via ultracold two-body collisions in the $s$-wave regime, providing a microscopic and tunable realization of the system--reservoir coupling.
Inelastic spin-exchange collisions induce internal state transitions within the hyperfine ground-state manifold of the atoms, and thereby mediate a quantized heat transfer between the working medium and the atomic reservoir.
Depending on the internal Zeeman state of the atomic bath, the spin-exchange collisions can be either exothermal or endothermal, allowing control over the direction of heat flow and, in turn, the realization of the heating and cooling steps of the quantum Otto cycle~\cite{kos17}.
The adiabatic expansion and compression strokes are further implemented via linear magnetic field ramps that modify the discrete level spacing of the working medium without changing its populations.

A central feature of our experimental system is that exothermal and endothermal spin-exchange collisions exhibit distinct energy scaling of their scattering cross sections.
As a result, the corresponding heat transfer dynamics between the working medium and the reservoir responds differently to changes in the collision energy distribution, which is set by the kinetic temperature of the atomic spin bath. 
This asymmetry allows us to control the relative durations of the isochoric heating and cooling strokes, and thereby the total cycle duration.
We exploit this feature to optimize the power output of the quantum heat engine at fixed efficiency by tuning the kinetic temperature of the atomic spin bath.

\section{Experimental system}
\label{sec:exp-real}
\begin{figure*}[htbp]
\begin{center}
\includegraphics[width=17.8 cm]{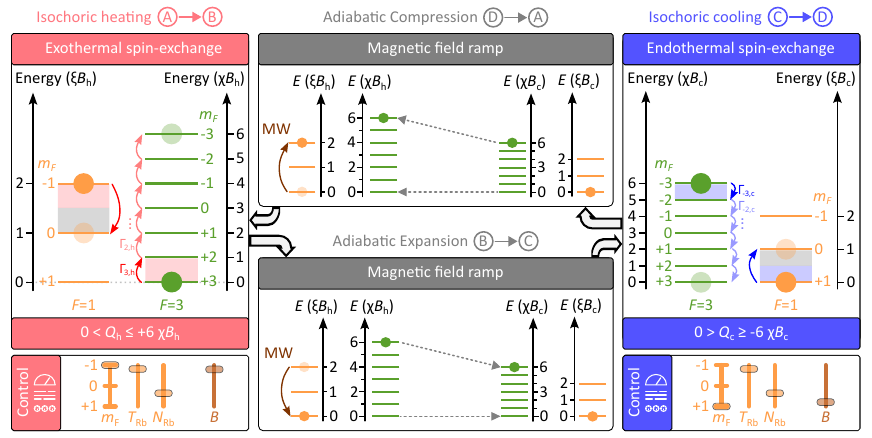}
\caption{{Operation scheme of the quantum Otto heat engine.} The working medium consists of the seven Zeeman states in the hyperfine ground-state manifold of Cs and undergoes a four-stroke cycle comprising two isochoric strokes and two adiabatic work strokes. The cycle starts with isochoric heating driven by exothermal spin-exchange collisions at a magnetic field $B_{\text{h}} = 1 \, \text{G}$, in which single quanta of heat (red shaded) are absorbed from the atomic reservoir of Rb atoms, while the excess internal energy (gray shaded) is released to the collision energy $\left[ \text{\Circled{A}} \to \text{\Circled{B}} \right]$. Microwave radiation then switches the internal Zeeman state of the atomic reservoir. The subsequent adiabatic expansion is realized by an adiabatic magnetic field ramp from $B_{\text{h}}$ to $B_{\text{c}} = 35 \, \text{mG}$, during which the working medium's populations do not change and work is performed $\left[ \text{\Circled{B}} \to \text{\Circled{C}} \right]$. The isochoric cooling is driven by endothermal spin-exchange collisions, transferring single quanta of heat (blue shaded) to the atomic reservoir, while the missing energy (gray shaded) is provided by the collision energy $\left[ \text{\Circled{C}} \to \text{\Circled{D}} \right]$. A final adiabatic field ramp from $B_{\text{c}}$ to $B_{\text{h}}$ completes the cycle $\left[ \text{\Circled{D}} \to \text{\Circled{A}} \right]$. The finite size of the working medium restricts the maximum number of inelastic spin-exchange collisions, and thereby also the number of transferred heat quanta, for each isochoric stroke to six. The state-dependent transition rates $\Gamma_{\text{h}}$ and $\Gamma_{\text{c}}$ are experimentally controlled via the atom number and the kinetic temperature of the atomic reservoir.}
\label{fig:cycle}
\end{center}
\end{figure*}

A schematic overview of the quantum Otto cycle in our experimental system is given in Fig.~\ref{fig:cycle}.
The basic building block of our quantum heat engine is a small, non-interacting ensemble of up to 40 neutral Cs atoms. 
These are immersed into a large, thermal cloud of optically trapped Rb atoms with fixed peak density $n_{\text{Rb},0} = 2.2(4) \times 10^{12} \, \text{cm}^{-3}$ and a tunable kinetic temperature $T_{\text{Rb}}$.
The measurements are performed at $T_{\text{Rb}} = \{679(175), 858(135), 1140(105)\} \, \text{nK}$, corresponding to total Rb atom numbers of \mbox{$N_{\text{Rb}} = \{ 3327(835), 4759(970), 8519(1655) \}$}, respectively~\footnote{Values and uncertainties are intentionally quoted with excess precision to provide reproducible input parameters for numerical simulations.}.
Details on the experimental sequence are given in App.~\ref{app:exp-methods}.
The internal hyperfine ground state of Cs acts as a quasi-spin with total angular momentum quantum number $F_{\text{Cs}} = 3$. 
The corresponding seven Zeeman states $m_{F,\text{Cs}} = +3, \dots, -3$ split linearly in the presence of a weak external magnetic field $B$ and represent the working medium of our quantum heat engine.
The energy of one of these Zeeman states is given by
\begin{equation}
	V_{\text{Cs}}(n, B) = (3 - m_{F,\text{Cs}}) \left| g_{F,\text{Cs}} \right| \mu_{\text{B}} B \equiv n \chi B,
	\label{eq:zeeman-energy}
\end{equation}
where we have set $n \equiv 3 - m_{F,\text{Cs}}$ and introduced the species-specific constant $\chi \equiv \left| g_{F,\text{Cs}} \right| \mu_{\text{B}}$ with Bohr magneton $\mu_{\text{B}}$ and Land\'{e} factor $g_{F,\text{Cs}} = -1/4$.
The zero-point of energy is set to the absolute ground state $m_{F,\text{Cs}} = +3$ of the Cs atoms.
The hyperfine ground-state manifold of Rb forms a quasi-spin with $F_{\text{Rb}} = 1$ and serves as the atomic reservoir, with all atoms initialized either in the $m_{F,\text{Rb}} = +1$ or $m_{F,\text{Rb}} = -1$ Zeeman state to set the direction of heat transfer for each isochoric stroke.
Analogously, the internal energy of the atomic bath is given by $V_{\text{Rb}}(\gamma, B) = (1 - m_{F,\text{Rb}}) \left| g_{F,\text{Rb}} \right| \mu_{\text{B}} B \equiv \gamma \xi B$ with \mbox{$\gamma = 1 - m_{F,\text{Rb}}$} and $\xi = \left| g_{F,\text{Rb}} \right| \mu_{\text{B}}$.
Importantly, the Land\'{e} factors of the two atomic species $g_{F,\text{Rb}} = -1/2 = 2g_{F,\text{Cs}}$ differ by a factor of two, i.e., the energy-level spacing of the Rb atoms is twice as large as the one of the Cs atoms ($\xi = 2 \chi$).

The Cs atoms and the Rb atoms interact via elastic and inelastic $s$-wave collisions~\cite{coh11}.
The system may thus be regarded as a physical realization of a so-called collisional model~\cite{Ciccarello2022}.
Frequent elastic collisions ensure the thermalization of the motional degree of freedom of the Cs atoms, and do not induce internal state transitions.
Hence, these do not contribute to the thermodynamic cycle and are thus irrelevant for its performance.
The less frequent inelastic collisions, by contrast, lead to state transitions within the hyperfine ground-state manifold
\begin{equation}
\begin{aligned}
	&\ket{1, m_{F,\text{Rb}}} \otimes \ket{3, m_{F,\text{Cs}}}\\ &\rightarrow \ket{1, m_{F,\text{Rb}} - \Delta m_F} \otimes \ket{3, m_{F,\text{Cs}} + \Delta m_F}, 
\end{aligned}
\end{equation}
where the total magnetization $M =  m_{F,\text{Rb}} + m_{F,\text{Cs}}$ is conserved and only transitions with $\Delta m_F = \pm 1$ are allowed.
Such inelastic spin-exchange collisions in the presence of a static magnetic field form the basis of the heat transfer between the working medium and the atomic reservoir.
Each spin-exchange collision is accompanied by a quantized energy transfer between the Cs atoms and the Rb cloud.
The induced population dynamics of the seven Zeeman states in the working medium directly reflects the heat exchange with the atomic reservoir
\begin{equation}
	Q(t) = \sum_{n=0}^6 V_{\text{Cs}}(n, B) \left[P_{n}(t) - P_{n}(t_0)\right],
	\label{eq:heat}
\end{equation} 
where $P_{n}(t_0)$ [$P_{n}(t)$] denotes the initial (final) population of the isochoric stroke.
As a result of the distinct Land\'{e} factors, internal energy $\abs{\Delta V} = (\xi - \chi)B$ is released to (absorbed from) the collision energy in a spin-exchange collision with $\Delta m_F = -1$ ($\Delta m_F = +1$) while altering the internal state of the atoms, making the collision exothermal (endothermal).

The adiabatic strokes in between the two isochoric strokes are realized by adiabatic magnetic field ramps that change only the level spacing of the working medium, thereby performing work
\begin{equation}
	W(t) = \sum_{n=0}^6 \left(V_{\text{Cs}}[n, B(t)] - V_{\text{Cs}}[n, B(t_0)]\right)P_n,
\end{equation}
with $B(t_0)$ [$B(t)$] being the magnetic field at the beginning (end) of the work stroke.

We employ time-resolved measurements of the working medium's population across the Otto cycle to extract the performance of the quantum heat engine; see App.~\ref{app:exp-methods} for details.  
The total cycle duration $\tau_{\text{cyc}} = \tau_{\text{h}} + \tau_{\text{c}} + 2 \tau_{\text{B}}$ is determined by the durations $\tau_{\text{h}}$ and $\tau_{\text{c}}$ of the two isochoric strokes, and by the time duration $\tau_{\text{B}} = 10 \, \text{ms}$ of each of the two adiabatic work strokes. 
The work strokes are sufficiently short that no spin-exchange collisions occur during them.
In particular, their duration is negligible compared to that of the isochoric strokes, which are on the order of a few seconds.
Therefore, the following power optimization is restricted to optimizing the durations of the isochoric strokes.
\vspace*{-5pt}
\section{Heat transfer law}
\label{sec:heat-trans-law}
In this section, we characterize the heat transfer law governing our system and elucidate its dependence on the kinetic temperature of the atomic bath. 
This will allow us to manipulate the microscopic heat flow. 
In contrast to the coupling to standard thermal reservoirs, heat transfer is here not determined by the temperature difference between system and bath. 
Its unusual properties therefore require a detailed analysis.
Heat transfer between the working medium and the atomic spin bath is instead mediated by inelastic spin-exchange collisions, which can be either exothermal (isochoric heating) or endothermal (isochoric cooling), as described in the previous section. 
These collisions induce state-dependent transitions between neighboring Zeeman states of the Cs hyperfine ground-state manifold and drive a directed population flow during the two isochoric strokes.
The resulting dynamics of the working medium can be described theoretically by a discrete master equation~\cite{Schmidt2018},
\begin{equation}
	\dot{\mathbf{P}} = K_i(T_{\text{Rb}}, B_i) \, \mathbf{P},
\end{equation}
where $\mathbf{P} = \left(P_3, P_2, \dots, P_{-3} \right)^\text{T}$ denotes the population vector of the seven Zeeman states and $i \in \{\text{h}, \text{c}\}$ labels the type of the isochoric stroke.
The formal solution is 
\begin{equation}
	\mathbf{P}(t) = \exp \left[K_i(T_{\text{Rb}}, B_i) t\right] \mathbf{P}(t_0),
	\label{eq:pop-evolution}
\end{equation}
showing that the heat transfer is fully determined by the rate matrix $K_i(T_{\text{Rb}}, B_i)$, which contains state-dependent transition rates.

We begin by examining the properties of the transition rates.
The rate matrices for both isochoric strokes are triangular matrices describing nearest-neighbor population transfer between Zeeman states of the working medium.
Isochoric heating is described by a lower triangular matrix, 
\begin{equation}
[K_{\text{h}}(T_{\text{Rb}}, B_{\text{h}})]_{m, m^\prime} =
\begin{cases}
-\Gamma_{m,\mathrm{h}}, & m = m^\prime, m \neq -3, \\
\Gamma_{m+1,\mathrm{h}}, & m = m^\prime - 1, \\
0, & \text{otherwise},
\end{cases}
	\label{eq:rate-mat-h}
\end{equation}
while isochoric cooling is described by an upper triangular matrix,
\begin{equation}
[K_{\mathrm{c}}(T_{\text{Rb}}, B_{\mathrm{c}})]_{m, m^\prime} =
\begin{cases}
-\Gamma_{m,\mathrm{c}}, & m = m^\prime, m \neq +3, \\
\Gamma_{m-1,\mathrm{c}}, & m = m^\prime + 1, \\
0, & \text{otherwise},
\end{cases}
	\label{eq:rate-mat-c}
\end{equation}
with $m, m^\prime \in \{+3, \dots, -3 \}$.
The rate matrix $K_{\text{h}}(T_{\text{Rb}}, B_{\text{h}})$ therefore generates directed population flow toward energetically higher Zeeman states (i.e., states with smaller magnetic quantum number $m_{F,\text{Cs}}$), whereas $K_{\text{c}}(T_{\text{Rb}}, B_{\text{c}})$ drives the reverse process.

The state-dependent transition rates are given by~\cite{Schmidt2018}
\begin{equation}
	\Gamma_{m, i} (T_{\text{Rb}}, N_{\text{Rb}}, B_i) = \overline{n(\mathbf{r}, T_{\text{Rb}})} \, \overline{\sigma_{m, i}(E_{\text{col}}, B_i) v_{\text{col}}},
	\label{eq:scat-rate}
\end{equation}
where $\overline{n(\mathbf{r}, T_{\text{Rb}})} = \int \text{d}\mathbf{r} \, n_{\text{Rb}}(\mathbf{r}, T_{\text{Rb}}) n_{\text{Cs}}(\mathbf{r}, T_{\text{Rb}})$ denotes the density overlap of the two atomic species.
The remaining rate coefficient $\overline{\sigma_{m, i}(E_{\text{col}}, B_i) v_{\text{col}}}$ captures the microscopic collision properties through the $s$-wave scattering cross section $\sigma_{m, i}(E_{\text{col}}, B_i)$ and the relative collision velocity $v_{\text{col}}$.
The scattering cross sections are obtained from coupled-channel scattering calculations, which include the internal (hyperfine) structure of both atomic species and their molecular interaction potential.
They depend not only on the external magnetic field $B_i$, but also on the collision energy $E_{\text{col}} = \mu v_{\text{col}}^2/2$, where $\mu$ is the reduced mass of the collision pair.
The collision energies follow a Maxwell--Boltzmann distribution, $p(E_{\text{col}}, T_{\text{Rb}})$, which is fully determined by the kinetic temperature of the atomic reservoir~\cite{Cannoni2014}.
Consequently, the scattering rates given in Eq.~\eqref{eq:scat-rate} are ensemble-averaged quantities depending on the rate coefficient
\begin{equation}
	\begin{aligned}
	&\overline{\sigma_{m, i}(E_{\text{col}}, B_i) v_{\text{col}}}\\
	&= \int_0^\infty \sigma_{m, i}(E_{\text{col}}, B_i) \sqrt{\frac{2 E_{\text{col}}}{\mu}} p(E_{\text{col}}, T_{\text{Rb}}) \, \text{d}E_{\text{col}}.
	\end{aligned}
	\label{eq:rate-coeff}
\end{equation}

\begin{figure}[tbp]
\begin{center}
\includegraphics[width=8.9 cm]{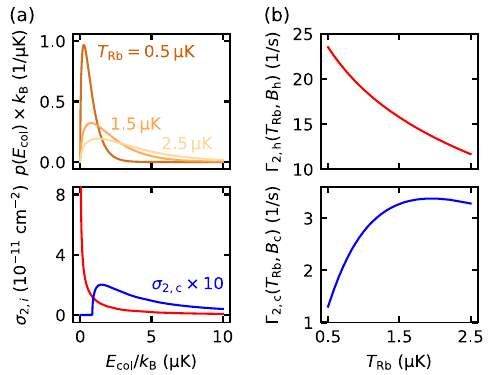}
\caption{{Inelastic $s$-wave scattering.} (a) Energy-dependent scattering cross sections of exothermal spin-exchange (red, $B_{\text{h}} = 1 \, \text{G}$) and endothermal spin-exchange \mbox{(blue, $B_{\text{c}} = 35 \, \text{mG}$)} for the Zeeman state $m_{F,\text{Cs}} = 2$. The kinetic temperature $T_{\text{Rb}}$ of the atomic reservoir sets the range of accessible collision energies via the distribution function $p(E_{\text{col}}, T_{\text{Rb}})$. (b) Transition rates resulting from the scattering cross sections shown in (a) by evaluating Eq.~\eqref{eq:scat-rate} for various kinetic temperatures $T_{\text{Rb}}$ of the atomic bath at a fixed peak density $n_{\text{Rb},0} = 2.2 \times 10^{12} \, \text{cm}^{-3}$.}
\label{fig:scat-prop}
\end{center}
\end{figure}

A key feature of the system is the distinct energy dependence of the $s$-wave scattering cross sections for exothermal and endothermal spin-exchange collisions.
The scattering cross sections $\sigma_{m,\text{h}}(E_{\text{col}}, B_{\text{h}})$ for exothermal spin-exchange decrease monotonically with collision energy, whereas the scattering cross sections $\sigma_{m,\text{c}}(E_{\text{col}}, B_{\text{c}})$ for endothermal spin-exchange exhibit a threshold behavior [Fig.~\ref{fig:scat-prop}(a)].
The magnitude of the corresponding rate coefficients is then set by the kinetic temperature of the atomic bath, which sets the range of accessible collision energies via the distribution function $p(E_{\text{col}}, T_{\text{Rb}})$ [Eq.~\eqref{eq:rate-coeff}]. 

Owing to the distinct energy scaling of the scattering cross sections, the transition rates $\Gamma_{m,\text{h}}(T_{\text{Rb}}, B_{\text{h}})$ and $\Gamma_{m,\text{c}}(T_{\text{Rb}}, B_{\text{c}})$ respond differently to the kinetic temperature of the atomic reservoir.
Specifically, the transition rates $\Gamma_{m,\text{h}}(T_{\text{Rb}}, B_{\text{h}})$ decrease monotonically with increasing kinetic temperature $T_{\text{Rb}}$, while the transition rates $\Gamma_{m,\text{c}}(T_{\text{Rb}}, B_{\text{c}})$ vary non-monotonically and attain a maximum [Fig.~\ref{fig:scat-prop}(b)].
This disparity in scaling induces an asymmetry between the two isochoric strokes: variations in $T_{\text{Rb}}$ affect the transition rates governing the heat transfer -- and thus the time required for each isochoric stroke -- in an inequivalent manner.
As a consequence, the relative time allocation between the isochoric heating and cooling strokes, and thereby the total cycle duration, becomes tunable via the kinetic temperature of the atomic reservoir.
The interplay between the monotonically decreasing transition rates $\Gamma_{m,\text{h}}(T_{\text{Rb}}, B_{\text{h}})$ and the non-monotonic transition rates $\Gamma_{m,\text{c}}(T_{\text{Rb}}, B_{\text{c}})$ implies that the total cycle duration is itself a non-monotonic function of $T_{\text{Rb}}$.
Since the output power is inversely related to the total cycle duration, it likewise exhibits a non-monotonic dependence on the kinetic temperature $T_{\text{Rb}}$, giving rise to an optimal kinetic temperature of the atomic reservoir, as we will see below.

\begin{figure}[htbp]
\begin{center}
\includegraphics[width=8.9 cm]{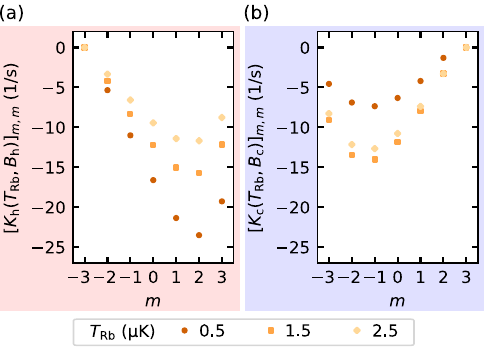}
\caption{{Eigenvalues of the rate matrices governing heat transfer during the two isochoric strokes.} Diagonal elements of the triangular rate matrices for isochoric heating (a) and isochoric cooling (b), shown for three kinetic temperatures of the atomic reservoir, $T_{\text{Rb}} = \{0.5, 1.5, 2.5 \} \, \mathrm{\text{\textmu} K}$, as obtained from Eqs.~\eqref{eq:rate-mat-h} and \eqref{eq:rate-mat-c}. 
Since the rate matrices are triangular, their diagonal elements coincide with the eigenvalues $\lambda_k$, which determine the relaxation timescales entering the multi-exponential heat transfer law in Eq.~\eqref{eq:heat-law}. 
The magnetic fields are set to $B_{\text{h}} = 1 \, \text{G}$ and $B_{\text{c}} = 35 \, \text{mG}$, and the peak density of the atomic reservoir is fixed at $n_{\text{Rb},0} = 2.2 \times 10^{12} \, \text{cm}^{-3}$.}
\label{fig:scat-prop-2}
\end{center}
\end{figure}

We next investigate the heat relaxation dynamics.
The heat transfer between the system and the bath is directly linked to the evolution of the population vector [Eq.~\eqref{eq:pop-evolution}] through
\begin{equation}
\begin{aligned}
	Q_i(t) &= \mathbf{V}^\text{T} [\mathbf{P}(t) - \mathbf{P}(t_0)] \\
	&= \sum_{k=0}^6 c_k d_k \left[\exp(\lambda_k t) - 1 \right],
\end{aligned}
\label{eq:heat-law}
\end{equation}
where $\mathbf{V}^\text{T} = (V_{\text{Cs}}[0, B_i], \dots, V_{\text{Cs}}[6, B_i)]$ denotes the vector of Zeeman energies and $d_k = \mathbf{V}^\text{T}\mathbf{v}_k$.
Here, the initial population vector has been decomposed into the eigenmodes of the rate matrix, $\mathbf{P}(t_0) = \sum_{k=0}^6 c_k \mathbf{v}_k$, where each eigenmode is a pair $(\lambda_k, \mathbf{v}_k)$ of eigenvalue and eigenvector.
The eigenvalues $\lambda_k$ of the triangular rate matrix are given by its diagonal elements $[K_{i}(T_{\text{Rb}}, B_i)]_{m, m}$, which are determined by the transition rates $\Gamma_{m, i}(T_{\text{Rb}}, B_i)$ [Eqs.~\eqref{eq:rate-mat-h} and \eqref{eq:rate-mat-c}].
Two types of modes arise: First, there is a stationary mode with $\lambda_k = 0$ corresponding to the fixed point of the dynamics.
Second, six relaxing modes with $\lambda_k < 0$ decay exponentially with relaxation times $\tau_k = -1/\lambda_k$.
Consequently, the heat transfer is governed by a weighted sum of exponential contributions associated with different relaxation modes.
The dynamics is hence not dominated by a single relaxing mode, as often assumed phenomenologically in macroscopic thermodynamics~\cite{Curzon1975, Andresen1984,gor91,and11,kos13,Chen1994_carnot,bro05,sch08,esp09,Esposito2010,whi14}. 
Instead, the spectral properties of the rate matrix exhibit a parabolic dependence on the Zeeman state, such that all non-zero eigenvalues have comparable magnitude (Fig.~\ref{fig:scat-prop-2}).
This distinguishes our system from usual classical relaxation processes, which typically display a mono-exponential relaxation, as in Newton's law of cooling~\cite{Curzon1975, Andresen1984,gor91,and11,kos13,Chen1994_carnot,bro05,sch08,esp09,Esposito2010,whi14}.

For short stroke durations, expanding the exponential in a Taylor series as $\exp(\lambda_k t) \approx 1 + \lambda_k t + \mathcal{O}(t^2)$ yields a linear heat response, $Q(t) \approx \left[\sum_{k=0}^6 c_k d_k \lambda_k \right]t + \mathcal{O}(t^2)$.
The slope of this initial linear growth corresponds to a weighted average of the eigenvalues of the relaxing modes, where $c_k$ quantifies how much of the initial population lies in that mode, and $d_k$ quantifies how energetically visible that mode is. 
For longer stroke durations, the dynamics enters a non-linear regime in which the exponential terms in Eq.~\eqref{eq:heat-law} separate the individual eigenmodes.
In the limit of infinitely long stroke durations, the exponentials vanish and the maximal transferable heat becomes $Q_{i,\infty} = - \sum_{k=0}^6 c_k d_k$.
Because of the finite size of the working medium and the directed population flow, these heats, $Q_{\text{h},\infty} = + 6 \chi B_{\text{h}}$ and $Q_{\text{c},\infty} = - 6 \chi B_{\text{c}}$, are determined solely by the maximal number of spin-exchange collisions and the external magnetic field, which sets the magnitude of the quantized energy exchanged with the atomic reservoir.

In order to quantify the non-exponential heat relaxation dynamics, we introduce the thermalization factor 
\begin{equation}
	G_i(t) = \frac{Q_i(t)}{Q_{i,\infty}} = \sum_{k=0}^6 w_k \left[1 - \exp (\lambda_k t) \right]
    \label{eq:therm-fac}
\end{equation}
as a measure of the fraction of equilibration, where we have introduced the new weights $w_k = c_k d_k / (\sum_{l=0}^6 c_l d_l)$.
Due to the normalization, the thermalization factor depends only on the system--reservoir interaction and satisfies $\lim \limits_{t \to \infty} G_i(t) = 1$, which corresponds to a full population inversion at the end of the isochoric stroke.

\begin{figure*}[tbp]
\begin{center}
\includegraphics[width=17.8 cm]{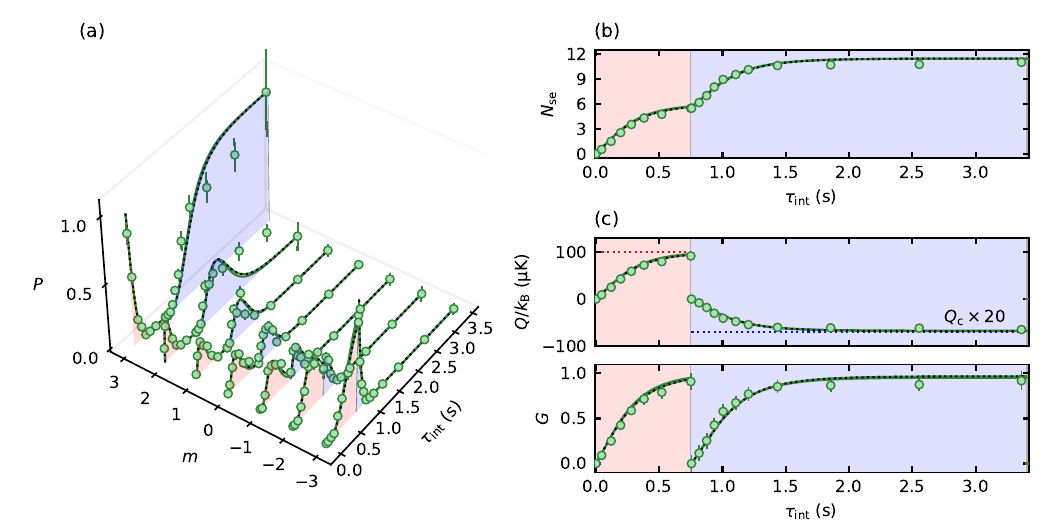}
\caption{{Heat transfer driven by inelastic spin-exchange collisions.} (a) Population dynamics of the seven Zeeman states in the working medium induced by exothermal spin-exchange collisions (red, isochoric heating) and endothermal spin-exchange collisions (blue, isochoric cooling). The working medium reaches full population inversion after an interaction time of approximately 750 ms. (b) Number-of-collisions evolution resulting from the measured spin dynamics. Full population inversion is reached after the maximal number of six spin-exchange collisions. (c) Heat transfer resulting from the spin-evolution driven by spin-exchange collisions. The evolution of the isochoric cooling stroke is magnified by a factor of 20. The working medium is nearly fully equilibrated at the end of each isochoric stroke, which is reflected by a thermalization factor approaching unity, as well as the transferred heat being close to its maximal possible value (red and blue dotted line). The kinetic temperature of the atomic reservoir is fixed to $T_{\text{Rb}} = 1140(105) \, \text{nK}$, and the magnetic fields are set to $B_{\text{h}} = 1 \, \text{G}$ and $B_{\text{c}} = 35 \, \text{mG}$. The solid lines show the working medium's dynamics resulting from our numerical model [Eq.~\eqref{eq:pop-evolution}] without introducing free parameters, and the black dotted lines show the prediction including also the finite lifetime of the Rb cloud. At this temperature, both predictions are nearly indistinguishable on the scale of the plots; the influence of the finite Rb bath lifetime becomes more pronounced at lower kinetic temperatures. The error bars are extracted from the statistical uncertainties in the atom number determination via standard error propagation.}
\label{fig:SE-3D}
\end{center}
\end{figure*}
Figure~\ref{fig:SE-3D} illustrates a representative measurement of the working medium's dynamics during heat transfer.
Inelastic spin-exchange collisions generate a directed population flow across the seven Zeeman states in the working medium, where the direction of heat flow is determined by the internal Zeeman state of the atomic reservoir~[Fig.~\ref{fig:SE-3D}(a)].
The corresponding number-of-collisions evolution reflects the sequential nature of the spin-exchange collisions during isochoric heating and cooling~[Fig.~\ref{fig:SE-3D}(b)].
The total number of spin-exchange collisions in each isochoric stroke is ultimately bounded by the finite size of the working medium, i.e., by the number of accessible Zeeman states, which amounts to six. 
The resulting heat transfer directly follows from the measured population dynamics and exhibits the characteristic multi-exponential behavior predicted by Eq.~\eqref{eq:heat-law}, with a clear separation between the initial approximately linear regime and the subsequent saturation toward the maximal transferable heat [Fig.~\ref{fig:SE-3D}(c)].

We may hence conclude that heat transfer in our system -- driven by directed exothermal or endothermal spin-exchange collisions -- generally follows a multi-exponential relaxation law with an approximate linear response at short times.
The equilibration dynamics is governed by state-dependent transition rates that originate from energy-dependent $s$-wave scattering cross sections.
Importantly, the distinct energy dependence of exothermal and endothermal scattering processes leads to qualitatively different scaling of the corresponding transition rates with the kinetic temperature of the atomic bath.
While the transition rates governing the isochoric heating decrease with increasing $T_{\text{Rb}}$, the rates corresponding to the isochoric cooling exhibit a non-monotonic behavior.
This asymmetry enables experimental control of the relative durations of the two isochoric strokes and makes the total cycle duration a non-monotonic function of the kinetic temperature $T_{\text{Rb}}$.

\section{Efficiency and power output}
\label{sec:merit-figures}
We proceed by introducing three key thermodynamic figures of merit that characterize the performance of our quantum heat engine:
efficiency $\eta$, power output $P$, and power output fluctuations $\sigma_P$.
All these quantities depend on the work output of the machine.
According to the first law of thermodynamics, we have
\begin{equation}
\begin{aligned}
	\abs{W(\tau_{\text{h}}, \tau_{\text{c}}, B_{\text{h}}, B_{\text{c}})} &= Q_{\text{h}}(\tau_{\text{h}}) - \abs{Q_{\text{c}}(\tau_{\text{c}})}\\
	&=6 \chi \left[B_{\text{h}} G_{\text{h}}(\tau_{\text{h}}) - B_{\text{c}} G_{\text{c}}(\tau_{\text{c}}) \right], 
\end{aligned}
\label{eq:work-output}
\end{equation}
where we have used Eq.~\eqref{eq:heat-law} for the heat $Q_i$ and Eq.~\eqref{eq:therm-fac} for the thermalization factor. 
To ensure cyclic operation, the working medium must return to its initial state after each cycle. 
This requires equal thermalization factors for the two isochoric strokes, $G_{\text{h}}(\tau_{\text{h}}) = G_{\text{c}}(\tau_{\text{c}})$.
Thus, the durations $\tau_{\text{h}}$ and $\tau_{\text{c}}$ are linked and cannot be chosen independently. 

The efficiency of the endoreversible quantum engine follows as the ratio of the work output and the heat input 
\begin{equation}
	\begin{aligned}
	\eta(B_{\text{h}}, B_{\text{c}}) = \frac{\abs{W(\tau_{\text{h}}, \tau_{\text{c}}, B_{\text{h}}, B_{\text{c}})}}{Q_{\text{h}}(\tau_{\text{h}}) + Q_{\text{l}}(\tau_{\text{h}}, \tau_{\text{c}})} 
	= \frac{\frac{\chi}{\xi}(B_{\text{h}} - B_{\text{c}})}{B_{\text{h}} - B_{\text{c}} + B_{\text{c}} \frac{\chi}{\xi}},
	\end{aligned}
	\label{eq:efficiency}
\end{equation}
where the heat leak $Q_{\text{l}}$ accounts for irreversible losses associated with the different Land\'{e} factors of Cs and Rb atoms~\cite{Bouton2021}.
The efficiency is only determined by the magnetic fields of the two isochoric strokes and the ratio $\chi/\xi = \left| g_{F,\text{Cs}} \right| / \left| g_{F,\text{Rb}} \right| = 0.5$ of the Land\'{e} factors of the two atomic species.
It is independent of the durations $\tau_{\text{h}}$ and $\tau_{\text{c}}$ of the two isochoric strokes, and reduces to the well-known reversible Otto efficiency $\eta_{\text{rev}} = 1 - B_{\text{c}}/B_{\text{h}}$~\cite{kos17} in the absence of irreversible losses, i.e., for $\chi/\xi = 1$.
On the other hand, the power output
\begin{equation}
\begin{aligned}
	P(\tau_{\text{h}}, \tau_{\text{c}}, B_{\text{h}}, B_{\text{c}}) &= \frac{\abs{W(\tau_{\text{h}}, \tau_{\text{c}}, B_{\text{h}}, B_{\text{c}})}}{\tau_{\text{cyc}} } \\
	&= \frac{6 \chi \left[B_{\text{h}} G_{\text{h}}(\tau_{\text{h}}) - B_{\text{c}} G_{\text{c}}(\tau_{\text{c}}) \right]}{\tau_{\text{h}} + \tau_{\text{c}} + 2 \tau_{\text{B}}}, 
\end{aligned}
\end{equation}
which characterizes the work production rate, and its fluctuations $\sigma_P = \sigma_{\abs{W}}/\tau_{\text{cyc}}$, determined from the population distribution of the seven Zeeman sublevels \cite{Bouton2021}, depend not only on the work output but also on the total cycle duration $\tau_{\text{cyc}} = \tau_{\text{h}} + \tau_{\text{c}} + 2 \tau_{\text{B}}$.
Thus, maximizing power output requires balancing two competing effects: increasing the extracted work by allowing sufficient equilibration and minimizing the total cycle duration.

The performance of the quantum heat engine can, in principle, be controlled via the parameters $\{ \tau_{\text{h}}, \tau_{\text{c}}, B_{\text{h}}, B_{\text{c}}\}$.
However, not all of these parameters provide equally effective control over the power output.
The magnetic fields $B_{\text{h}}$ and $B_{\text{c}}$ primarily determine the energy turnover per cycle.
In particular, the work output scales linearly with the magnetic field difference $B_{\text{h}} - B_{\text{c}}$~[Eq.~\eqref{eq:work-output}].
At the same time, the magnetic fields also enter the transition rates in Eq.~\eqref{eq:scat-rate} and thereby influence the dynamics of the heat transfer.
In the experimentally relevant parameter regime, however, this influence on the transition rates, and thus on the optimal stroke durations, is comparatively weak; see App.~\ref{app:b-variation} for further details.
As a result, variations in $B_{\text{h}}$ and $B_{\text{c}}$ predominantly lead to a rescaling of the work output, while corresponding changes in the optimal cycle duration remain negligible. 
Consequently, any gain in output power obtained by modifying the magnetic fields is dominated by the trivial linear increase of the work output, rather than by a genuine optimization of the engine dynamics.

In contrast, the time durations $\tau_{\text{h}}$ and $\tau_{\text{c}}$ directly control the trade-off between work output and total cycle duration, and therefore constitute relevant parameters for the power optimization.
In our system, these durations are not tuned independently, but are set by the equilibration dynamics during the isochoric strokes, which is governed by the kinetic temperature of the atomic reservoir.
The distinct response of the transition rates for exothermal and endothermal spin-exchange collisions to $T_{\text{Rb}}$ provides the possibility of selectively tuning the relative durations of the two isochoric strokes via the kinetic temperature of the atomic reservoir, as discussed in the previous section. 
This enables us to experimentally control the time allocation within the cycle and thereby optimize the power output at fixed magnetic fields.

\section{Power enhancement}
\label{sec:results}
This section presents the experimental results on the power optimization of the engine with respect to the isochoric stroke durations, which are controlled via the kinetic temperature of the atomic reservoir; the magnetic fields $B_{\text{h}} = 1 \, \text{G}$ and $B_{\text{c}} = 35 \, \text{mG}$ are kept constant.
We measure the performance for three distinct kinetic temperatures of the atomic reservoir, \mbox{$T_{\text{Rb}} = \{679(175), 858(135), 1140(105)\} \, \text{nK}$}, at a fixed peak density $n_{\text{Rb},0} = 2.2(4) \times 10^{12} \, \text{cm}^{-3}$.
The figures of merit are extracted from measurements of the population evolution of the working medium during heat transfer.
We compare the experimental data with our numerical model, which predicts the dynamics of the working medium without introducing any free parameters.

\begin{figure}[htbp]
\begin{center}
\includegraphics[width=8.9 cm]{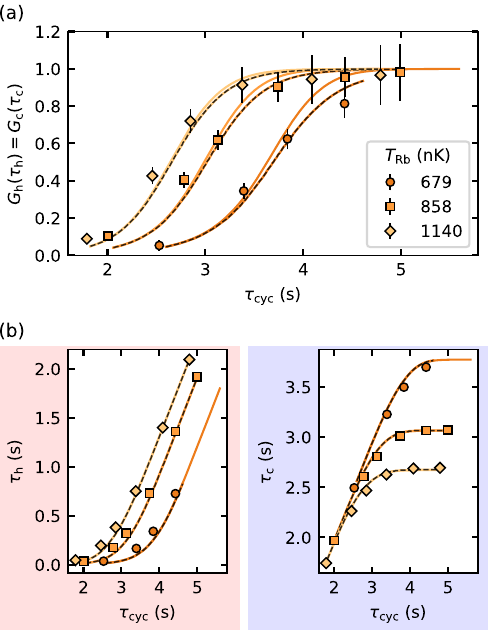}
\caption{{Controlling the equilibration dynamics via the kinetic temperature of the atomic reservoir.} (a) Evolution of the thermalization factor $G_{\text{h}}(\tau_{\text{h}}) = G_{\text{c}}(\tau_{\text{c}})$ as a function of the total cycle duration $\tau_{\text{cyc}}$ for different kinetic temperatures $T_{\text{Rb}}$ of the atomic reservoir. (b) Stroke durations required to reach a fixed fraction of equilibration, illustrating the redistribution of time allocation within the cycle. Increasing $T_{\text{Rb}}$ induces a pronounced asymmetry: isochoric cooling accelerates, while isochoric heating slows down. The solid lines show the equilibration dynamics resulting from our numerical model [Eq.~\eqref{eq:pop-evolution}] without introducing free parameters, and the black dashed lines show the prediction including also the finite lifetime of the Rb cloud. The error bars originate from the statistical uncertainties in the atom number determination via standard error propagation.}
\label{fig:therm-fac}
\end{center}
\end{figure}

We first analyze the equilibration dynamics, which determines the time allocation within the cycle and thereby sets the basis for power optimization.
Figure~\ref{fig:therm-fac} illustrates the equilibration dynamics through the thermalization factor [Eq.~\eqref{eq:therm-fac}] as a function of the total cycle duration $\tau_{\text{cyc}} = \tau_{\text{h}} + \tau_{\text{c}} + 2 \tau_{\text{B}}$.
The evolution of the thermalization factor shows that increasing the kinetic temperature of the atomic reservoir accelerates the overall equilibration dynamics within the experimentally explored parameter regime: for a given fraction of equilibration, $G_{\text{h}}(\tau_{\text{h}}) = G_{\text{c}}(\tau_{\text{c}})$ -- and thus for a fixed work output $\abs{W(\tau_{\text{h}}, \tau_{\text{c}}, B_{\text{h}}, B_{\text{c}})}$ according \mbox{to Eq.~\eqref{eq:work-output} -- the} corresponding total cycle duration is reduced at higher kinetic temperatures $T_{\text{Rb}}$ [Fig.~\ref{fig:therm-fac}(a)].
However, the whole equilibration dynamics is not uniformly accelerated.
Instead, increasing $T_{\text{Rb}}$ induces an asymmetric modification of the timescales of the two isochoric strokes~[Fig.~\ref{fig:therm-fac}(b)]: the isochoric heating stroke slows down due to decreasing exothermal transition rates $\Gamma_{m,\text{h}}(T_{\text{Rb}}, B_{\text{h}})$, while the isochoric cooling stroke accelerates as the endothermal transition rates $\Gamma_{m,\text{c}}(T_{\text{Rb}}, B_{\text{c}})$ increase with the kinetic temperature $T_{\text{Rb}}$ within the experimentally explored regime~[Fig.~\ref{fig:scat-prop}(b)]. 
As a result, the relative time allocation within the cycle is systematically shifted, leading to a reduction of the total cycle duration with increasing kinetic temperature of the atomic reservoir.
These observations demonstrate that the time allocation is governed by the energy dependence of the underlying inelastic scattering processes and can be tuned via the kinetic temperature of the atomic reservoir.
This establishes a link between microscopic scattering dynamics and macroscopic engine performance, thereby providing a control mechanism for power optimization: 
reducing the total cycle duration for a given fraction of equilibration $G_{\text{h}}(\tau_{\text{h}}) = G_{\text{c}}(\tau_{\text{c}})$ directly enhances the power output, which is inversely proportional to $\tau_{\text{cyc}}$.
Furthermore, we observe that the measured equilibration dynamics is systematically slower than the dynamics predicted by our numerical model, which is particularly apparent for the lowest employed kinetic temperature $T_{\text{Rb}} = 679(175) \, \text{nK}$~[Fig.~\ref{fig:therm-fac}(a), solid lines].
This systematic deviation can be attributed to the finite lifetime of the Rb atoms in the optical dipole trap, which manifests in a continuous heating of the Rb sample and Rb atom loss from the trap, and affects cold clouds particularly strongly; measurements of the Rb lifetime are summarized in App.~\ref{app:rb-lifetime}. 
Although an increasing kinetic temperature $T_{\text{Rb}}$ would by itself suggest faster equilibration, the accompanying reduction of the peak density of the Rb bath due to cloud expansion and atom loss lowers the microscopic transition rates governing heat transfer.
This effect dominates under the present experimental conditions and results in an overall slowdown of the equilibration dynamics.
We account for this effect by adding a time-dependent kinetic temperature and density of the atomic reservoir to our numerical model, yielding good agreement with the experimental data [Fig.~\ref{fig:therm-fac}(a), black dashed lines]. 

\begin{figure*}[tbp]
\begin{center}
\includegraphics[width=17.8 cm]{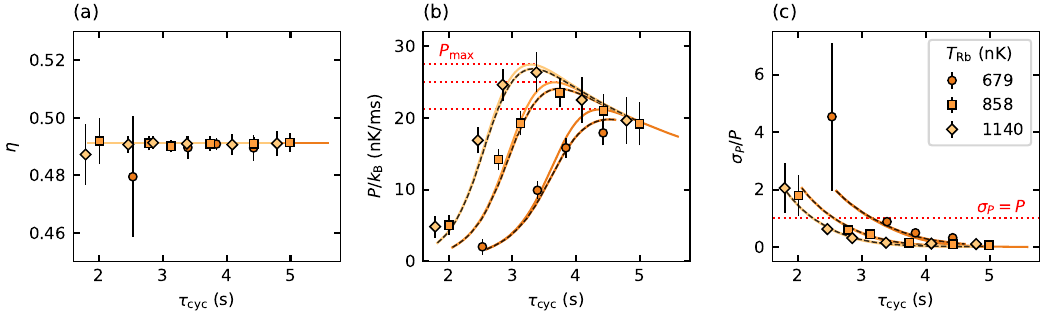}
\caption{{Controlling the performance by the kinetic temperature of the atomic reservoir.} (a) Efficiency $\eta$, (b) power output $P$, and (c) relative power fluctuations $\sigma_P/P$ as a function of the total cycle duration $\tau_{\text{cyc}}$ for different kinetic temperatures $T_{\text{Rb}}$ of the atomic reservoir. The efficiency remains constant, independent of both $\tau_{\text{cyc}}$ and $T_{\text{Rb}}$, in agreement with the endoreversible model. By contrast, the output power exhibits a non-trivial dependence on $\tau_{\text{cyc}}$, with a maximum that shifts with $T_{\text{Rb}}$. This behavior reflects the equilibration dynamics governed by state-dependent scattering rates, which exhibit a distinct energy dependence for isochoric heating and cooling strokes.
The relative power fluctuations decrease with increasing total cycle duration, and the maximum power output is consistently reached within the low-fluctuation regime. The solid lines show the performance evolution predicted by our numerical model [Eq.~\eqref{eq:pop-evolution}] without introducing free parameters, whereas the black dashed lines [(b) and (c)] show the performance taking also the finite lifetime of the Rb cloud into account. The error bars are extracted from the statistical uncertainties in the atom number determination via standard error propagation.}
\label{fig:performance}
\end{center}
\end{figure*}

Having established how the kinetic temperature of the atomic reservoir controls the equilibration dynamics and total cycle duration, we now turn to its impact on the engine performance.
We evaluate the performance for different total cycle durations $\tau_{\text{cyc}}$ by varying the stroke durations $\tau_{\text{h}}$ and $\tau_{\text{c}}$, and extract the transferred heat from the measured population dynamics of the working medium.
Figure~\ref{fig:performance} shows how the kinetic temperature $T_{\text{Rb}}$ modifies the efficiency $\eta$, the power output $P$, and the relative power output fluctuations $\sigma_P/P$. 
First, we observe a constant efficiency $\eta = 0.489(2)$ independent of the kinetic temperature and the total cycle duration, in agreement with the expected value $\eta(B_{\text{h}}, B_{\text{c}}) = 0.491$ from Eq.~\eqref{eq:efficiency} [Fig.~\ref{fig:performance}(a)].
Second, the power output $P$ exhibits a non-trivial dependence on $\tau_{\text{cyc}}$, with a maximum $P_{\text{max}}$ that increases with $T_{\text{Rb}}$ within the experimentally explored parameter range [Fig.~\ref{fig:performance}(b)].
This behavior directly reflects the modified equilibration dynamics discussed above.
Third, we quantify the stability of the quantum heat engine via the relative power fluctuations $\sigma_P/P$ [Fig.~\ref{fig:performance}(c)]~\cite{pie18,hol18,den21}. 
The engine transitions from a strongly fluctuating regime ($\sigma_P/P > 1$) at short total cycle durations to a stable regime with sub-Poissonian fluctuations ($\sigma_P/P < 1$) at longer durations.
Importantly, for all kinetic temperatures, the maximum power is consistently reached within this low-fluctuation regime, demonstrating that optimal performance is achieved without sacrificing either efficiency or stability.
Overall, we find again good quantitative agreement between the experimental data and our numerical model.
 
We now combine the insights from the equilibration dynamics (Fig.~\ref{fig:therm-fac}) and the performance evolution (Fig.~\ref{fig:performance}) to identify and interpret the optimal operating point of the quantum heat engine.
The optimal total cycle duration emerges from the multi-exponential nature of the heat transfer: since no single characteristic timescale governs the equilibration, the optimal stroke durations result from the interplay of multiple relaxation modes and cannot be inferred from a simple rate argument, i.e., there is no closed-form analytic solution for the optimal stroke durations.
\begin{figure}[htbp]
\begin{center}
\includegraphics[width=8.9 cm]{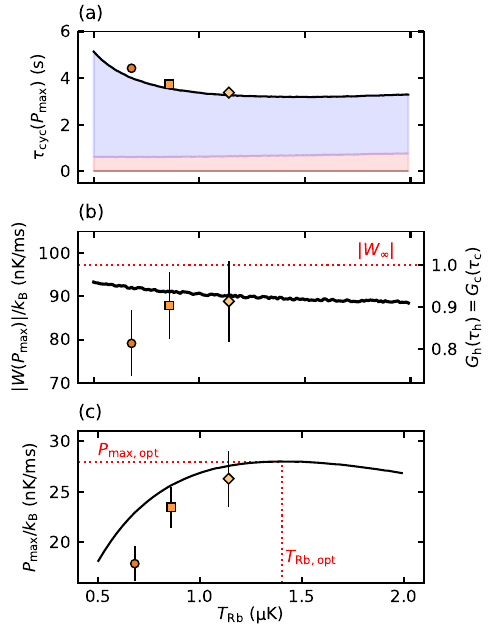}
\caption{{Optimal operating point of the quantum heat engine.} (a) Optimal total cycle duration $\tau_{\text{cyc}}(P_{\text{max}})$ versus the kinetic temperature $T_{\text{Rb}}$ of the atomic reservoir. The non-monotonic dependence arises from the asymmetric scaling of the transition rates $\Gamma_{m,\text{h}}(T_{\text{Rb}}, B_{\text{h}})$ (isochoric heating, red) and $\Gamma_{m,\text{c}}(T_{\text{Rb}}, B_{\text{c}})$ (isochoric cooling, blue), which modifies the time allocation within the cycle. (b) Work output $\abs{W(P_{\text{max}})}$ and thermalization factor $G_{\text{h}}(\tau_{\text{h}}) = G_{\text{c}}(\tau_{\text{c}})$ at maximum power, showing that optimal performance is achieved close to full equilibration with weak dependence on $T_{\text{Rb}}$. (c) Maximum power output $P_{\text{max}}$ versus $T_{\text{Rb}}$, exhibiting a well-defined optimum $P_{\text{max},\text{opt}}$ at finite kinetic temperature $T_{\text{Rb},\text{opt}}$. Together, these results demonstrate that power optimization is governed by the interplay between multi-exponential equilibration dynamics and the asymmetric energy dependence of the transition rates governing heat transfer, enabling control of the total cycle duration via $T_{\text{Rb}}$. The solid lines show the prediction from our numerical model [Eq.~\eqref{eq:pop-evolution}] without introducing free parameters.}
\label{fig:performance-opt}
\end{center}
\end{figure}
Figure~\ref{fig:performance-opt}(a) shows that the optimal total cycle duration $\tau_{\text{cyc}}(P_{\text{max}})$ depends non-monotonically on the kinetic temperature $T_{\text{Rb}}$.
This behavior directly reflects the asymmetric temperature dependence of the transition rates governing the isochoric heating and cooling strokes, which modifies the relative time allocation within the cycle.
The comparison with the equilibration dynamics reveals that maximum power is achieved for a thermalization factor close to unity at low kinetic temperatures, indicating that the working medium operates near full population inversion and that the extracted work $W(P_{\text{max}})$ is close to its maximal value $W_{\infty} = 6\chi [B_{\text{h}} - B_{\text{c}}]$ [Fig.~\ref{fig:performance-opt}(b)]. 
With increasing kinetic temperature, the thermalization factor decreases monotonically, indicating increasingly incomplete equilibration during the isochoric strokes and a corresponding reduction of the work turnover per cycle.
The resulting maximum power output therefore exhibits a non-monotonic dependence on $T_{\text{Rb}}$ [Fig.~\ref{fig:performance-opt}(c)].
This non-monotonic behavior arises from the competition of two opposing effects: 
at low kinetic temperatures, the power output is primarily limited by the long total cycle durations required to achieve near-complete equilibration.
Increasing $T_{\text{Rb}}$ accelerates the heat transfer dynamics and thereby reduces the total cycle duration, entering a regime in which the reduced total cycle duration overcompensates the gradual reduction in work turnover per cycle and leads to a higher maximum power output.
At higher kinetic temperatures, however, the thermalization factor continues to decrease while the total cycle duration reaches a minimum and subsequently increases slightly.
Consequently, the reduction in work turnover becomes the dominant effect, causing the maximum power to decrease again with $T_{\text{Rb}}$.
As a result, the quantum heat engine exhibits an optimal operating point at finite kinetic temperature $T_{\text{Rb,opt}} = 1410(5) \, \text{nK}$, characterized by a peak power of $P_{\text{max},\text{opt}}/k_{\text{B}} = 28.0 \, \text{nK}/\text{ms}$.

\begin{figure}[htbp]
\begin{center}
\includegraphics[width=8.9 cm]{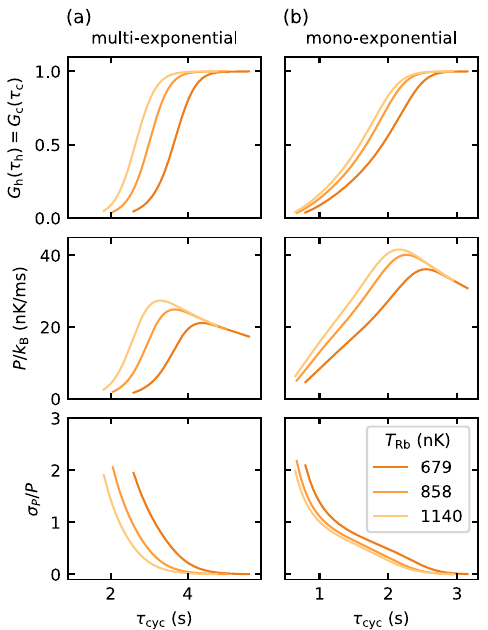}
\caption{{Comparison of multi-exponential and mono-exponential heat transfer dynamics.} Evolution of the thermalization factor $G_{\text{h}}(\tau_{\text{h}}) = G_{\text{c}}(\tau_{\text{c}})$, the power output $P$, and the relative power fluctuations $\sigma_P/P$ as a function of the total cycle duration $\tau_{\text{cyc}}$ for different kinetic temperatures $T_{\text{Rb}}$ of the atomic reservoir.
(a) Full model with state-dependent transition rates, giving rise to multi-exponential equilibration. (b) Reference model with state-independent transition rates obtained by averaging the state-dependent scattering rates, resulting in mono-exponential equilibration. Compared with the multi-exponential case, the mono-exponential dynamics reaches the same degree of equilibration in shorter cycle times, reducing the optimal cycle duration by approximately a factor of $0.56$--$0.62$ and increasing the corresponding maximum power by approximately a factor of $1.52$--$1.71$.}
\label{fig:comparison}
\end{center}
\end{figure}

To assess the impact of the multi-exponential heat transfer dynamics on the engine performance relative to the usually assumed phenomenological single-exponential relaxation~\cite{Curzon1975, Andresen1984,gor91,and11,kos13,Chen1994_carnot,bro05,sch08,esp09,Esposito2010,whi14}, we numerically compare our model to a reference model with mono-exponential heat-exchange.
We construct the reference model by replacing the state-dependent transition rates $\Gamma_{m, i} (T_{\text{Rb}}, N_{\text{Rb}}, B_i)$~[Eq.~\eqref{eq:scat-rate}] with the corresponding average transition rate $\Gamma_{i} (T_{\text{Rb}}, N_{\text{Rb}}, B_i) = \sum_m \Gamma_{m, i} (T_{\text{Rb}}, N_{\text{Rb}}, B_i) / 6$, where $i \in \{\text{h}, \text{c}\}$.
This replacement removes the distribution of relaxation times while preserving the average transition rate for each isochoric stroke.
Figure~\ref{fig:comparison} compares the resulting equilibration dynamics and engine performance with those of the full multi-exponential model.
We observe a faster thermalization than in the microscopic multi-exponential case, which leads to a higher power output, while the relative power fluctuations remain comparable. 
The single-exponential relaxation assumption hence overestimates the performance of the engine with respect to the actual multi-exponential heat dynamics. 
Such a single-exponential assumption should therefore be used with caution.
\section{Conclusion and Outlook}
\label{sec:conclusion-outlook}
We have demonstrated that experimental control of microscopic system-reservoir interactions enables both the manipulation of the heat exchange at the nanoscale and the optimization of the finite-time performance of a quantum heat engine.
By exploiting the asymmetric energy dependence of inelastic scattering processes, we have controlled the time allocation within the thermodynamic cycle and thereby maximized power output.
Beyond the specific realization of a collision-driven quantum Otto heat engine, these results highlight a more general principle: the finite-time performance of a thermal machine can be optimized through microscopic control of the system--reservoir interaction itself~\cite{pan20}, as opposed to the control of the system Hamiltonian, as done so far~\cite{ben17,mye22,arr23,can24}.
In most experimental and theoretical studies of quantum heat engines, heat transfer is furthermore described phenomenologically in terms of effective thermalization rates or dissipative couplings~\cite{Curzon1975, Andresen1984,gor91,and11,kos13,Chen1994_carnot,bro05,sch08,esp09,Esposito2010,whi14}.
By contrast, in our system, these rates emerge directly from state-resolved inelastic scattering properties whose energy dependence can be engineered through the collision energy distribution.
This establishes a direct link between microscopic collision processes and macroscopic thermodynamic performance. 

From a broader perspective, our experiment provides a physical implementation of a collision-model paradigm, in which equilibration arises from repeated interactions with a system and individual constituents of a reservoir~\cite{Ciccarello2022}.
Such collision-based descriptions have become a versatile framework for open quantum systems, non-equilibrium thermodynamics, and reservoir engineering~\cite{Ziman2005,Scarani2002,Karevski2009}. 
Our results establish that modifying the microscopic interaction law can enhance the performance of quantum engines, complementary to more commonly explored strategies based on quantum coherence, nonthermal reservoirs, and engineered bath correlations~\cite{Manzano2018,Cakmak2019,Watanabe2017}.

Importantly, the underlying mechanism and hence our findings are not restricted to ultracold gases. 
Similar state-dependent relaxation processes may arise in a variety of systems, including central-spin models, spin-bath environments, and solid-state quantum impurities coupled to surrounding spin degrees of freedom~\cite{Anderson1954,Yang2017,Prokofev2000}. 
In all these systems, the relevant control parameter is not the thermodynamic state of the reservoir alone, but the structure of the relaxation spectrum governing the energy exchange.
Concretely, the coexistence of multiple relaxation channels naturally leads to non-exponential equilibration dynamics and nontrivial system--bath energy exchange. 
Our results suggest that such microscopic relaxation mechanisms may be exploited as resources for controlling thermodynamic performance rather than merely constituting sources of dissipation.

Future work may extend these ideas to non-Markovian heat reservoirs or optimized control protocols beyond fixed-cycle operation.

\begin{acknowledgments}
We thank A. Guthmann for providing us with the scattering cross sections underlying our numerical model.
This work was supported by the Deutsche Forschungsgemeinschaft (DFG, German Research Foundation) via the Collaborative Research Center SFB/TR185 (Project No. 277625399), and has received funding from the European Research Council (ERC) under the European Union’s Horizon Europe research and innovation programme (Advanced Grant “QuantumEngine”, grant agreement No. 101200776).
S.B. acknowledges funding by the Studienstiftung des deutschen Volkes.
\end{acknowledgments}

\section*{\label{sec:data-availability}Data availability statement}
The data that support the findings of this study are openly
available at the following URL/DOI: \url{https://doi.org/10.5281/zenodo.21903512}.

\clearpage

\onecolumngrid
\appendix
\renewcommand{\thefigure}{A\arabic{figure}}
\setcounter{figure}{0}

\section{\label{app:exp-methods}Experimental sequence and observables}
This section gives an overview of the experimental methods and the typical observables in our experimental system.
The initial preparation of the two atomic samples takes place in spatially separated regions of our experimental setup.

We begin with the all-optical preparation of a thermal Rb cloud in the magnetically insensitive state $\ket{1, 0}$ of the $5S_{1/2}$ electronic ground state.
Standard absorption imaging after $7 \, \text{ms}$ of time-of-flight is used to determine the final atom number $N_{\text{Rb}}$ and temperature $T_{\text{Rb}}$ of the Rb sample.

In the next step, a small ensemble of Cs atoms is prepared in the $\ket{3, 3}$ state of the $6S_{1/2}$ electronic ground state via laser cooling followed by degenerate Raman sideband cooling.
The Cs atoms are transferred to a second crossed optical dipole trap located at an axial distance of approximately $200 \, \mathrm{\text{\textmu m}}$ from the trap containing the thermal Rb cloud. 
Residual Cs atoms that do not occupy the absolute ground state $\ket{3, 3}$ are removed using a combination of microwave Landau-Zener sweeps and resonant laser pulses, based on the spin-selective readout technique described in Ref.~\cite{Schmidt2018}.
This procedure yields a spin-polarized Cs ensemble of approximately $N_{\text{Cs}} \simeq 40$ Cs atoms in the absolute ground state.

In the final preparation step, a species-selective, one-dimensional optical lattice formed by two counter-propagating $790 \, \text{nm}$ laser beams is used to transport the Cs atoms into the Rb cloud~\cite{Schmidt2016}.
The lattice is subsequently switched off for the duration of the quantum heat engine operation. 

The quantum Otto cycle is initiated after the Rb cloud is transferred to the $\ket{1, -1}$ state via two successive microwave Landau-Zener sweeps with a total duration of $8.4 \, \text{ms}$.
During the cycle, the internal state of the Rb atoms is alternated between $\ket{1, -1}$ and $\ket{1, +1}$ using successive microwave Landau-Zener sweeps. 
The total duration of each state transfer amounts to $10.6 \, \text{ms}$, which is much shorter than the characteristic timescale of the spin-exchange interaction.

The magnetic fields $B_{\text{h}} = 1 \, \text{G}$ and $B_{\text{c}} = 35 \, \text{mG}$ are calibrated using Rb microwave spectroscopy~\cite{Bouton2021}.
The linear magnetic field ramps of the work strokes must be adiabatic, such that the populations of the seven Zeeman states of the Cs atoms remain unchanged.
This requirement can be expressed by the adiabaticity criterion $\abs{\dot{\omega}_{\text{L}}} \ll \omega_{\text{L}}^2(t)$, where $\omega_{\text{L}}(t) = \chi B(t)/\hbar$
denotes the time-dependent Larmor frequency, with $\hbar$ being the reduced Planck constant.
Consequently, adiabaticity is ensured for ramp durations satisfying $\tau_{\text{B}} \gg \hbar \abs{B_{\text{c}} - B_{\text{h}}} / (\chi B_{\text{c}}^2) = 0.36 \, \text{ms}$.
In the experiment, the magnetic field is ramped linearly between $B_{\text{h}}$ and $B_{\text{c}}$ (or vice versa) within $\tau_{\text{B}} = 10 \, \text{ms}$, thereby fulfilling the adiabaticity condition at all times.
Moreover, $\tau_{\text{B}}$ remains much shorter than the characteristic spin-exchange timescale, ensuring that Zeeman state populations of the Cs atoms remain unchanged during the work strokes.

Importantly, the species-selective optical lattice is not only employed for transport but also enables spatially resolved fluorescence imaging of the Cs atoms after their interaction with the Rb cloud.
The lattice creates a repulsive potential for the Cs atoms, thereby pinning their positions along the lattice axis while the Rb cloud is removed from the optical dipole trap using a resonant laser pulse. 
Subsequently, spatially resolved fluorescence imaging is combined with a sequence of microwave Landau-Zener sweeps and resonant laser pulses to measure the spin distribution of the Cs atoms~\cite{Schmidt2018}.
\section{\label{app:b-variation}Heat transfer at various magnetic fields}
The aim of this section is to substantiate the choice of the isochoric stroke durations $\tau_{\text{h}}$ and $\tau_{\text{c}}$ as the primary control parameters for power optimization, while keeping the magnetic fields $B_{\text{h}}$ and $B_{\text{c}}$ fixed. 

\begin{figure*}[!t]
\begin{center}
\includegraphics[width=17.8 cm]{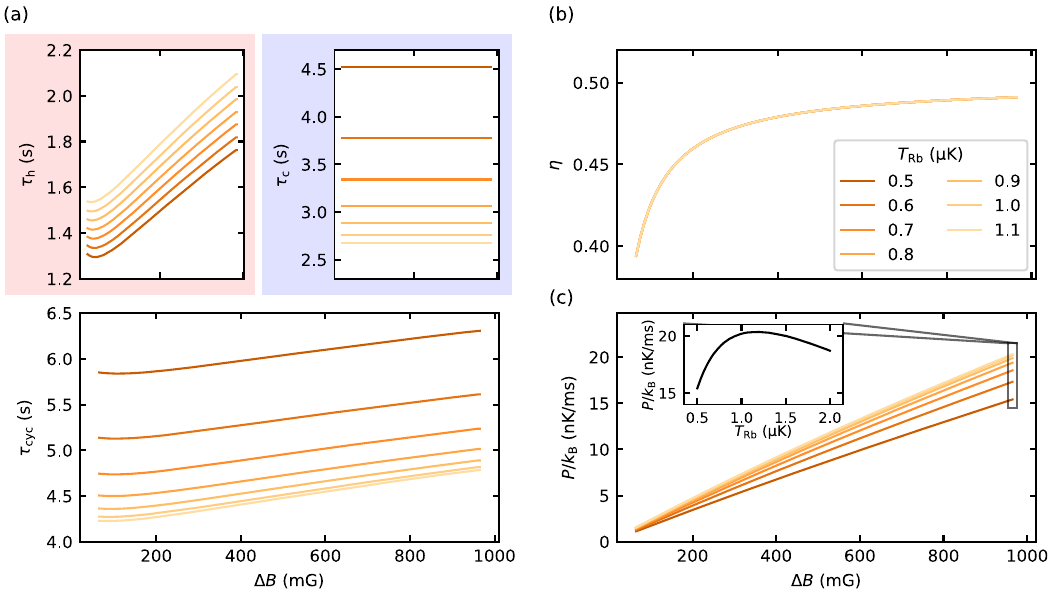}
\caption{{Heat transfer dynamics and performance at varying magnetic fields.} (a) Stroke durations $\tau_{\text{h}}$ and $\tau_{\text{c}}$ required to reach near-complete population inversion ($G_{\text{h}}(\tau_{\text{h}}) = G_{\text{c}}(\tau_{\text{c}}) \approx 1$), as well as the resulting total cycle duration $\tau_{\text{cyc}}$, as functions of the kinetic temperature $T_{\text{Rb}}$ and the magnetic field difference $\Delta B = B_{\text{h}} - B_{\text{c}}$. While $\tau_{\text{cyc}}$ increases weakly with $\Delta B$ due to the magnetic field dependence of the heating dynamics, its variation is dominated by the kinetic temperature $T_{\text{Rb}}$. (b) Efficiency $\eta$ and (c) power output $P$ of the quantum heat engine as a function of $\Delta B$.
The efficiency increases monotonically with $\Delta B$, in agreement with Eq.~\eqref{eq:efficiency}. The power output exhibits an approximately linear scaling with $\Delta B$, reflecting the increase in work output per cycle, while changes in the total cycle duration remain subdominant. These results show that varying the magnetic fields predominantly rescales the energy turnover per cycle, whereas the cycle dynamics, and thus the optimal time allocation, are primarily controlled by the kinetic temperature $T_{\text{Rb}}$ of the atomic reservoir. 
}
\label{fig:performance-pop-inv}
\end{center}
\end{figure*}

To this end, we numerically evaluate the performance of the quantum heat engine for a fixed thermalization factor $G_{\text{h}}(\tau_{\text{h}}) = G_{\text{c}}(\tau_{\text{c}}) \approx 1$, corresponding to (near) complete population inversion at the end of the isochoric heating stroke. 
The corresponding stroke durations $\tau_{\text{h}}$ and $\tau_{\text{c}}$ are extracted from the population dynamics of the working medium by imposing that the target state $m_{F,\text{Cs}} = -3$ (isochoric heating) and $m_{F,\text{Cs}} = +3$ (isochoric cooling) are populated with $99.5 \%$ probability.
Figure~\ref{fig:performance-pop-inv}(a) shows the resulting stroke durations and the corresponding total cycle duration $\tau_{\text{cyc}}$ as functions of the kinetic temperature $T_{\text{Rb}}$ and the magnetic field difference $\Delta B = B_{\text{h}} - B_{\text{c}}$, within the experimentally relevant parameter regime.
In this analysis, the magnetic field during the isochoric cooling stroke is kept fixed at $B_{\text{c}} = 35 \, \text{mG}$, such that the cooling duration $\tau_{\text{c}}$ depends only on the kinetic temperature $T_{\text{Rb}}$ and is independent of $\Delta B$.
By contrast, variations in $\Delta B$ directly affect the isochoric heating stroke via the magnetic field $B_{\text{h}}$, which enters the state-dependent scattering rates [Eq.~\eqref{eq:scat-rate}].
As a consequence, the total cycle duration $\tau_{\text{cyc}}$ exhibits a weak, approximately linear increase with $\Delta B$, originating from the magnetic field dependence of the isochoric heating dynamics.
However, this dependence is weaker than the variation induced by the kinetic temperature.
Across the considered parameter range, $\tau_{\text{cyc}}$ increases by at most a factor of $1.1$ when varying $\Delta B$, whereas it changes by up to a factor of $1.4$ when varying $T_{\text{Rb}}$.

We now turn to the performance of the quantum heat engine.
The efficiency $\eta$, given by Eq.~\eqref{eq:efficiency}, is independent of the stroke durations $\tau_{\text{h}}$ and $\tau_{\text{c}}$ as well as the kinetic temperature $T_{\text{Rb}}$, but increases monotonically with the magnetic field difference $\Delta B$ [Fig.~\ref{fig:performance-pop-inv}(b)].
The power output $P$ also depends on $\Delta B$, primarily through the work output, which scales linearly with the magnetic field difference [Fig.~\ref{fig:performance-pop-inv}(c)].
Although $\tau_{\text{cyc}}$ increases slightly with $\Delta B$, this effect remains subdominant.
We therefore conclude that varying the magnetic fields does modify the power output, primarily through the linear scaling of the work per cycle. 
However, this change does not arise from an optimization of the underlying finite-time dynamics, but rather from a trivial rescaling of the energy turnover per cycle. 
In particular, the location of the optimal operating point and the characteristic cycle times remain essentially unaffected.

By contrast, the stroke durations $\tau_{\text{h}}$ and $\tau_{\text{c}}$, which are governed by the equilibration dynamics, directly control the interplay between work extraction and total cycle duration. 
These durations can be tuned via the kinetic temperature $T_{\text{Rb}}$ of the atomic reservoir, providing a genuine handle to optimize the power output through dynamical control.
This justifies the restriction adopted in the main text to fixed magnetic fields, $B_{\text{h}} = 1 \, \text{G}$ and $B_{\text{c}} = 35 \, \text{mG}$, and optimization solely with respect to the stroke durations, $\tau_{\text{h}}$ and $\tau_{\text{c}}$.

\section{\label{app:rb-lifetime}Rb lifetime}
We quantify the lifetime of the Rb cloud in the optical dipole trap by measuring the atom number $N_{\text{Rb}}$ and the kinetic temperature $T_{\text{Rb}}$ after a time-of-flight of $7 \, \text{ms}$, as a function of the duration $\tau_j$ of the isochoric stroke, with $j \in \{\text{h}, \text{c} \}$. 

A phenomenological description for the time evolution of the Rb cloud parameters is given by
\begin{equation}
	T_{\text{Rb}}(\tau_j) = T_{\text{Rb}, \text{ss}} - ( T_{\text{Rb}, \text{ss}} -  T_{\text{Rb}, 0}) \exp \left(- \frac{\gamma_{\text{ht}} \tau_j}{T_{\text{Rb}, \text{ss}}} \right)
	\label{eq:fit-T-rb}
\end{equation}
\begin{equation}
	N_{\text{Rb}}(\tau_j) = N_{\text{Rb}, 0} - \gamma_{\text{ev}} \tau_j,
	\label{eq:fit-N-rb}
\end{equation}
which allows us to fit the experimental data using the free parameters $T_{\text{Rb}, \text{ss}}$, $T_{\text{Rb}, 0}$, $\gamma_{\text{ht}}$, $N_{\text{Rb}, 0}$ and $\gamma_{\text{ev}}$.
This, in turn, enables us to incorporate the finite lifetime of the Rb cloud into our numerical model.

\begin{figure*}[!t]
\begin{center}
\includegraphics[width=17.8 cm]{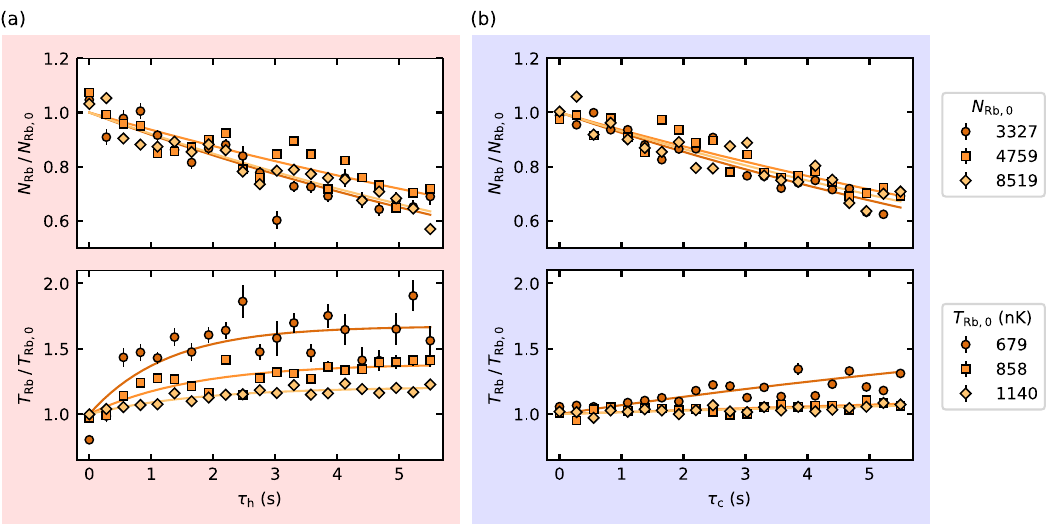}
\caption{{Rb cloud lifetime.} Lifetime measurement of a thermal Rb cloud at fixed peak density $n_{\text{Rb},0} = 2.2(4) \times 10^{12} \, \text{cm}^{-3}$ and variable kinetic temperature $T_{\text{Rb}}$ for (a) the isochoric heating stroke ($m_{F,\text{Rb}} = -1$) and (b) the isochoric cooling stroke ($m_{F,\text{Rb}} = +1$). The solid lines show fits according to Eqs.~\eqref{eq:fit-T-rb} and \eqref{eq:fit-N-rb}. Error bars arise from the fitting uncertainty of the measured Rb line density.}
\label{fig:rb-lifetime}
\end{center}
\end{figure*}

Figure~\ref{fig:rb-lifetime} shows that the Rb cloud undergoes continuous heating during the operation of the quantum heat engine, an effect that is particularly pronounced for initially cold clouds.
The heating is accompanied by a steady loss of Rb atoms from the optical dipole trap, which appears to be independent of both the kinetic temperature and the internal Zeeman state of the cloud.

We attribute the finite lifetime of the Rb cloud to processes such as recombination heating due to three-body collisions, technical noise, and spontaneous scattering of trap photons~\cite{Weber2003, Gehm1998, Grimm1999}.

\clearpage

\twocolumngrid
\bibliography{bibliography}

\end{document}